\documentclass[aps,prb,twocolumn,superscriptaddress,float,amsmath]{revtex4-2}
\usepackage{xcolor}
\usepackage{graphicx}
\usepackage{amsmath,bm,epsfig}
\definecolor{g-blue}{rgb}{0.83,0.95,1}
\usepackage{bm}
\usepackage{amssymb}
\usepackage{amsfonts}
\usepackage{amsmath}
\usepackage{graphicx}
\usepackage{ulem}
\usepackage{upgreek} 
\usepackage{textcomp} 
\usepackage[pdftex,colorlinks=true,allcolors=blue,breaklinks=true]{hyperref}  

\begin{document}
	
	\title{Towards an experimental proof of the magnonic Aharonov--Casher effect}
	 
	\author{Rostyslav~O.~Serha}
	\email{rostyslav.serha@univie.ac.at}
	\affiliation{Fachbereich Physik and Landesforschungszentrum OPTIMAS, Rheinland-Pf\"{a}lzische Technische Universit\"{a}t Kaiserslautern-Landau, 67663 Kaiserslautern, Germany}
    \affiliation{University of Vienna, Faculty of Physics, Vienna 1090, Austria}
	
	\author{Vitaliy~I.~Vasyuchka}
	\email{vitaliy.vasyuchka@rptu.de}
	\affiliation{Fachbereich Physik and Landesforschungszentrum OPTIMAS, Rheinland-Pf\"{a}lzische Technische Universit\"{a}t Kaiserslautern-Landau, 67663 Kaiserslautern, Germany}

	
	\author{Alexander~A.~Serga}
	\email{serha@rptu.de}
	\affiliation{Fachbereich Physik and Landesforschungszentrum OPTIMAS, Rheinland-Pf\"{a}lzische Technische Universit\"{a}t Kaiserslautern-Landau, 67663 Kaiserslautern, Germany}

	\author{Burkard~Hillebrands}
	\email{hillebra@rptu.de}
	\affiliation{Fachbereich Physik and Landesforschungszentrum OPTIMAS, Rheinland-Pf\"{a}lzische Technische Universit\"{a}t Kaiserslautern-Landau, 67663 Kaiserslautern, Germany}

	
	\begin{abstract}
Controlling the phase and amplitude of spin waves in magnetic insulators with an electric field opens the way to fast logic circuits with ultra-low power consumption. One way to achieve such control is to manipulate the magnetization of the medium via magnetoelectric effects. In experiments with magnetostatic spin waves in an yttrium iron garnet film, we have obtained the first evidence of a  theoretically predicted phenomenon: The change of the spin-wave phase due to the magnonic Aharonov--Casher effect---the geometric accumulation of the magnon phase as these quasiparticles propagate through an electric field region.
	\end{abstract}
	
	
	\maketitle
 
Magnonics, a research branch in the field of magnetism \cite{Kruglyak2010, Lenk2011, Chumak2015, Pirro2021}, offers a promising chance to overcome the challenges of modern complementary metal-oxide-semiconductor (CMOS) technology \cite{Barman2021}. Unlike conventional electronics, which uses voltage-controlled charge currents to process data, magnonics employs magnons, bosonic quasiparticles of collective spin oscillations---spin waves \cite{Pirro2021}. Magnons transport energy and magnetic moment in the form of a charge-free spin current \cite{Maekawa2023} without Joule heating and with wavelengths down to the nanoscale \cite{Che2020, Dieterle2019}. The latter allows us to imagine a strong miniaturization of future magnon-based data processing circuits \cite{Heinz2022}. These collective excitations in magnetically ordered materials provide conventional wave properties such as interference, as well as the quantum nature properties of quasiparticles with nonlinearity and Bose--Einstein condensation capability \cite{Demokritov2006, Serga2014, Schneider2020}, making them ideal data carriers for future computing concepts \cite{Chumak2022, Dieny2020, Mohseni2022, Csaba2017, Khitun2010}.
	
Magnonics enables wave logic, which exploits the superposition of coherent waves involving linear and nonlinear effects. Therefore, phase control of spin waves is an important issue.
Furthermore, to transport and process information \cite{Mohseni2022}, it has been proposed to use macroscopic magnon quantum phenomena, such as the room-temperature magnon Bose--Einstein condensate (BEC) and magnon supercurrents, which are phase-induced collective motions of quantum condensates \cite{Bozhko2019}. This approach requires precise control of the magnon BEC phase.
	
Already, success has been achieved in controlling the spin-wave phase using various magnetoelectric (ME) effects \cite{Fiebig2005, Matsukura2015, Hirosawa2022}. Methods to excite \cite{Cherepov2014, Vanderveken2020} and modulate \cite{Balinskiy2018} spin waves via the ME effects, based on magnetostriction and piezoelectricity, were elaborated. Another intensively investigated ME effect is the voltage-controlled magnetic anisotropy, which influences the surface layer properties of ferromagnets \cite{Miwa2019, Endo2010, Weisheit2007}. This phenomenon was already used to excite spin waves and oscillations \cite{Verba2014, Nozaki2012, Zhu2012} and to shift their resonance frequency \cite{Ha2010}. The possibility of a ME-controlled frequency and phase shift of a spin wave propagating in a magnetic insulator, namely, in yttrium iron garnet (Y$_3$Fe$_5$O$_{12}$, YIG), a material that is widely used in magnonics due to its lowest known magnetic losses \cite{Cherepanov1993, Serga2010, Pirro2021}, was analyzed theoretically \cite{Wang2018} and realized experimentally \cite{Zhang2014, Ustinov2019, Yu2019}.
	
Another approach, motivating from both a purely scientific and a practical point of view, is using the Aharonov--Casher (AC) effect \cite{Aharonov1984, Liu2011, Nakata2017, Krivoruchko2018_2, Savchenko2019, Krivoruchko2020}. This effect lies in the geometric accumulation of the wave function's phase as particles with magnetic moments pass through an electric field region:
\begin{equation}
    \Delta \varphi_{\mathrm{AC}} = \frac{1}{\hbar c^2} \int_{P} \mathbf{E}\times\mathbf{\upmu}\;\textrm{d}\mathbf{r}\,.
\end{equation}
Here, $\Delta \varphi_{\mathrm{AC}}$, $\mathbf{E}$, $c$ and $\mathbf{\upmu}$ are the AC phase, the external electric field, the speed of light and the magnetic moment of a (quasi-) particle moving along the path $P$ \cite{Nakata2017}. 

This phenomenon has been observed for real particles \cite{Cimmino1989} and theoretically predicted for quasiparticles such as magnons \cite{Cao1997}. Moreover, the AC effect is believed to be applied even to magnons with zero group velocity. Thus, it can be used to create and control persistent currents of a magnon BEC \cite{Nakata2015, Nakata2014}. 

However, the previous studies \cite{Zhang2014} on manipulating the magnon phase in YIG using an electric field did not allow one to distinguish the AC effect from the contribution of ME effects. It is related to the expectedly small influence of the AC effect on the long-wave magnetostatic spin waves studied in the experiment \cite{Liu2011}. \looseness=-1

Here, we report experimental results in favor of the existence of the mag\-no\-nic Aharonov--Casher effect. These results were obtained in an experiment with surface and backward volume magnetostatic spin waves propagating in an in-plane magnetized YIG film. Using two types of magnetostatic waves propagating perpendicular and parallel to the film magnetization direction, and thus differently affected by the AC effect, allowed us to reveal a contribution to the electric field-induced phase shift, which we attribute to the AC phase.
	
\begin{figure*}  
	\includegraphics[width=1.0\linewidth]{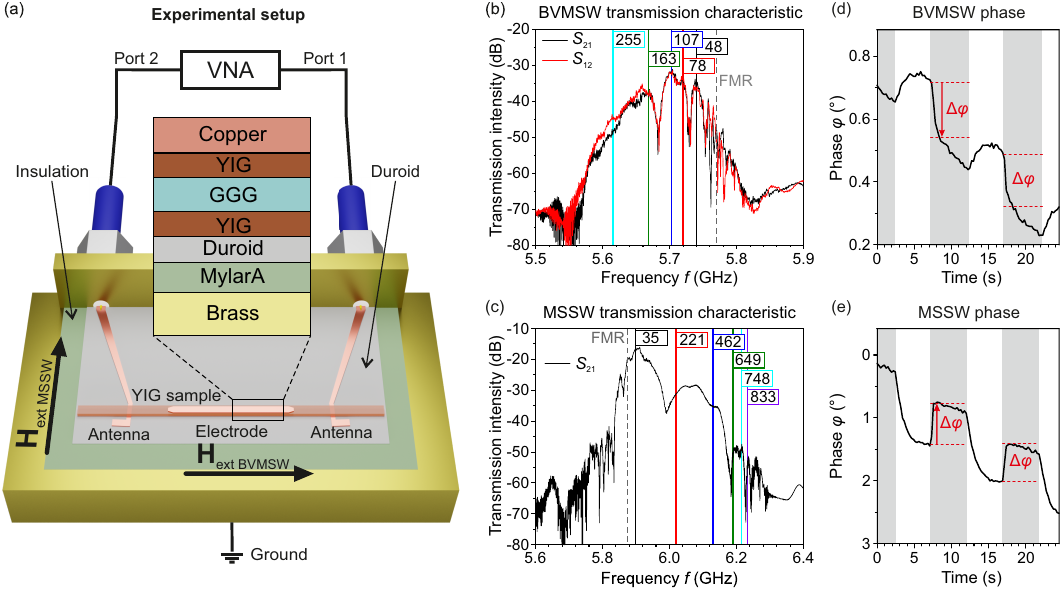} 
        \caption{(a) Sketch of the experimental setup. The YIG sample is placed on copper microstrip antennas and exposed to the external magnetizing field $H_\mathrm{ext}$. Spin waves are excited and measured using a vector network analyzer (VNA). By applying a voltage onto the copper electrode on the upper sample surface, an electric field is created between the electrode and the grounded brass sample holder. The inset illustrates the multilayer structure between these electrodes. (b) and (c) present the transmission characteristics measured for the BVMSW and MSSW geometries with the highlighted frequencies and estimated wavevector values (shown in rad/cm on the labels) selected for the phase shift measurement. The expected value of the ferromagnetic resonance frequency at zero wavevector is shown by a dashed gray line. The measurements were performed at $H_\mathrm{ext} = 136.4$\,mT for BVMSW and 138.9\,mT for MSSW geometries. (d) and (e) show examples of time-phase measurements performed for the $S_{21}$ parameter in the BVMSW ($f=5.74$\,GHz, $k=78$\,rad/cm) and MSSW ($f=6.21$\,GHz, $k=748$\,rad/cm) geometries, respectively. The gray background indicates the time intervals when the electric field with a strength of 4\,kV/mm was turned on. The electrically-induced phase shifts $\Delta \varphi$ are highlighted by dashed horizontal red lines. The linear phase shift, regardless of whether the electric field is on or off, is a continuous phase drift caused by temperature instability.
		}
	\label{f:1}
\end{figure*}

The experimental setup is shown schematically in Fig.\,\ref{f:1}a. 
The YIG sample contains a 7.8\,$\upmu$m-thick YIG film grown by liquid-phase epitaxy on a 500\,$\upmu$m-thick gallium gadolinium garnet substrate in the (111) plane. It is cut as a 41\,mm-long and 2\,mm-wide strip, which is oriented along the $\langle 110 \rangle$ direction and acts as a waveguide for spin waves. 
In the experiment, the YIG strip was magnetized by an external magnetic field $H_\mathrm{ext}$ both along and perpendicular to its long axis to provide backward volume magnetostatic spin waves (BVMSW) and magnetostatic surface spin waves (MSSW) geometries, respectively.
To minimize phase jumps and drifts, this field was created by a permanent magnet located together with the setup in a temperature-stabilized environment. 

The YIG film faces a pair of 50\,$\upmu$m-wide microstrip spin-wave antennas spaced 16\,mm apart. The antennas are chemically etched on a 350\,$\upmu$m-thick plate of copper-laminated ceramic-PTFE composite Rogers RT/duroid\textsuperscript{\tiny\textregistered} with the dielectric constant $\varepsilon_\mathrm{r} = 10.2$. The spin wave excited by the microwave magnetic field of one of the antennas propagates through the YIG film to the second one, where it is received and measured. A vector network analyzer (VNA) Anritsu MS4642B was used as a microwave source and a precise phase-resolved microwave detector.
 
The inset in Fig.\,\ref{f:1}b shows the multilayer structure in the area where the electric field is applied. 
At the bottom is a grounded brass sample holder. On top of it is an insulating layer of 500\,$\upmu$m-thick MylarA\textsuperscript{\tiny\textregistered} (PET plastic). 
It is followed by the 320\,$\upmu$m-thick duroid\textsuperscript{\tiny\textregistered} layer carrying the microstrip antennas, and on top of that is a 516\,$\upmu$m thick YIG/GGG/YIG sample. Finally, a 20\,$\upmu$m-thick, 10\,mm-long, and 1.5\,mm-wide copper electrode is glued to the sample's surface and connected to the high-voltage source (Advanced Energy 60C24-P60-I5).
Considering the glue between the layers, the applied voltage creates an electric field at a distance of $1.34 \pm 0.005$\,mm. 
The electrode, sample, and antennas were coated with a thin superglue layer, allowing the safe application of electric fields of up to 6\,kV/mm.

First, for a given magnetic field $\mathbf{H}_\mathrm{ext}$, the intensity of the output signal was measured as a function of frequency, and several frequencies were selected for phase observations (see Figs.\,\ref{f:1}b and \ref{f:1}c). 

Next, the phase and amplitude of the output signal were measured using the VNA in quasi-oscilloscope mode at fixed frequencies for a large number of consecutive points in time.

In this operating mode, the electric voltage was switched on and off several times during the VNA sweep. The phase difference between the time intervals when the electric field was on and off (see  Figs.\,\ref{f:1}d and \ref{f:1}e) is considered as the electrically-induced phase shift $\Delta \varphi$. 
The temporal behavior of the phase switching is mainly determined by the characteristics of the high-voltage source.
	
\begin{figure}  
		\includegraphics[width=1.0\linewidth]{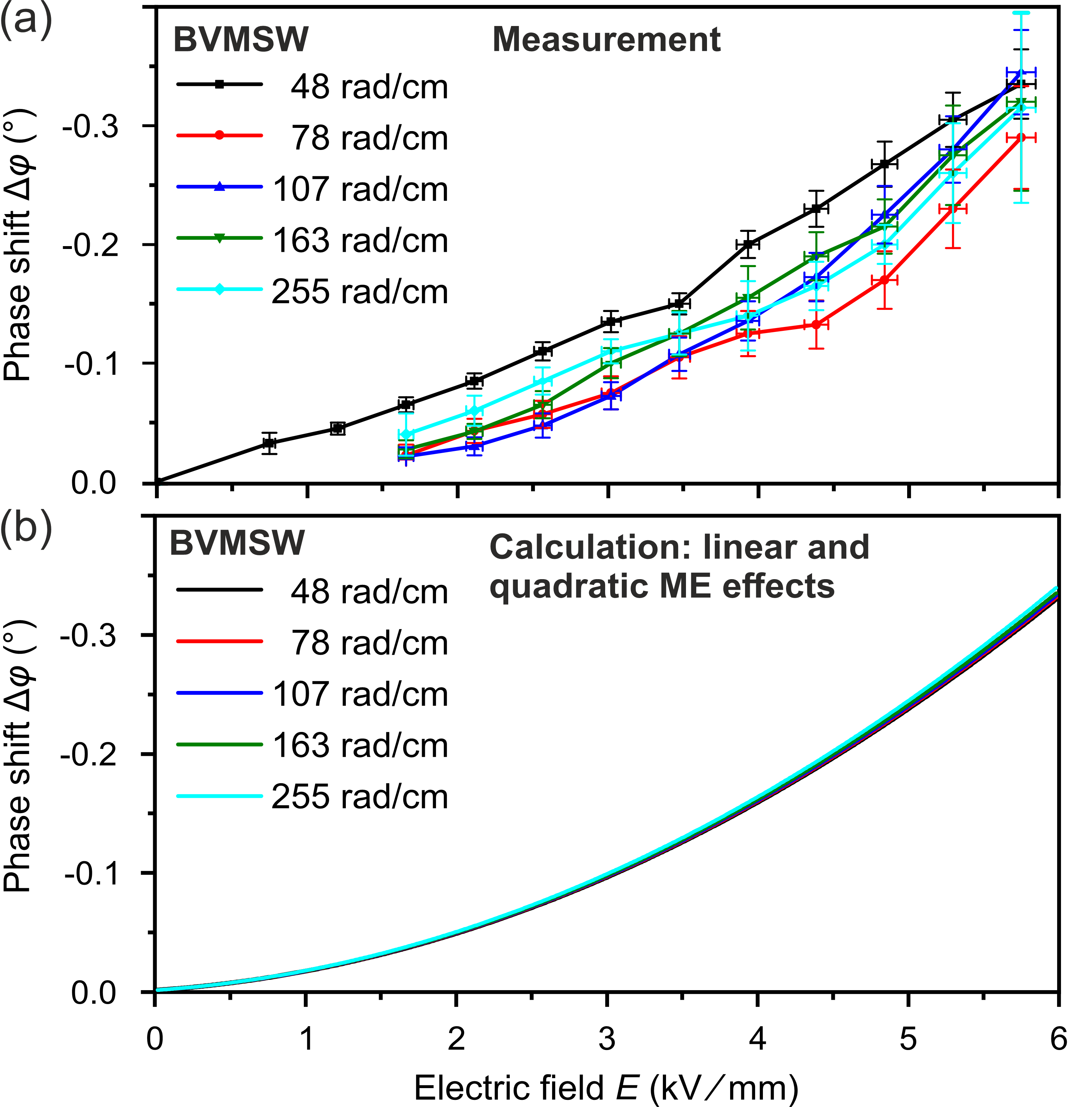}
		\caption{(a) Measured BVMSW phase shifts versus the applied electric field for five different wavevectors. (b) The calculated BVMSW phase shifts. The calculation was performed considering the contributions of the quadratic and linear ME effects. The corresponding ME coefficients were found by fitting the experimental data.
				}
		\label{f:2}
\end{figure}

Initially, the measurements were performed in the BVMSW geometry when magnons propagate along $\mathbf{H}_\mathrm{ext}$. When an electric field $\mathbf{E}$ is applied, a negative phase shift is observed, as shown in Fig.\,\ref{f:1}d.
The phase shifts of BVMSWs propagating in opposite directions are the same within the 95\% confidence interval. Their mean values are plotted in Fig.\,\ref{f:2}a as a function of the electric field strength $E = |\mathbf{E}|$.

The measurements were carried out at five different frequencies and, accordingly, different BVMSW wavevectors. Figure\,\ref{f:2}a shows that the phase shift is rather small ($\Delta \varphi \leq 0.35$\textdegree) and can be considered independent of the magnon wavevector. 
While the strongest shifts, shown by the black curve, are observed for the wavevector value $k = |\mathbf{k}| = 48$\,rad/cm, and the smallest, shown by the red curve, are for $k = 78$\,rad/cm, the other three dependencies for larger $k$ are in between. Moreover, these three curves are mostly within each other's error bars. From the shape of the curves in Fig.\,\ref{f:2}a, the phase shift seems to have a quadratic dependence on $E$.

The increase in measurement errors with increasing $k$ is due to a decrease in spin wave transmission, as is shown in Fig.\,\ref{f:1}b. This reduction limits the maximum available wavevector value to $k = 255$\,rad/cm. Phase noise also made it impossible to measure the phase shift for electric fields below $1.5$\,kV/mm, except for the strongest spin wave with the smallest wavevector.

The second series of measurements concerns MSSWs. In this geometry, the wavevector $\mathbf{k}$ is perpendicular to $\mathbf{H}_\mathrm{ext}$, and $\mathbf{E}$ is perpendicular to both. The phase-shift measurement example in Fig.\,\ref{f:1}e shows that $\Delta \varphi$ is positive in this case. The difference in the sign of the phase shifts of MSSW and BVMSW is due to the opposite directions of the group velocities of these two types of waves.
Since MSSWs are excited nonreciprocally \textcolor{red}{\cite{Serha2022}}, only $\Delta \varphi$ for one propagation direction can be measured. The corresponding transmission characteristic \textit{S}\textsubscript{12} is shown in Fig.\,\ref{f:1}c together with the six frequencies chosen for the measurements and the corresponding calculated values of the MSSW wavevector $k$. The broader bandwidth of the MSSW allows the use of much higher wavevectors for measurements than in the BWMSW case.

\begin{figure}  
		\includegraphics[width=1.0\linewidth]{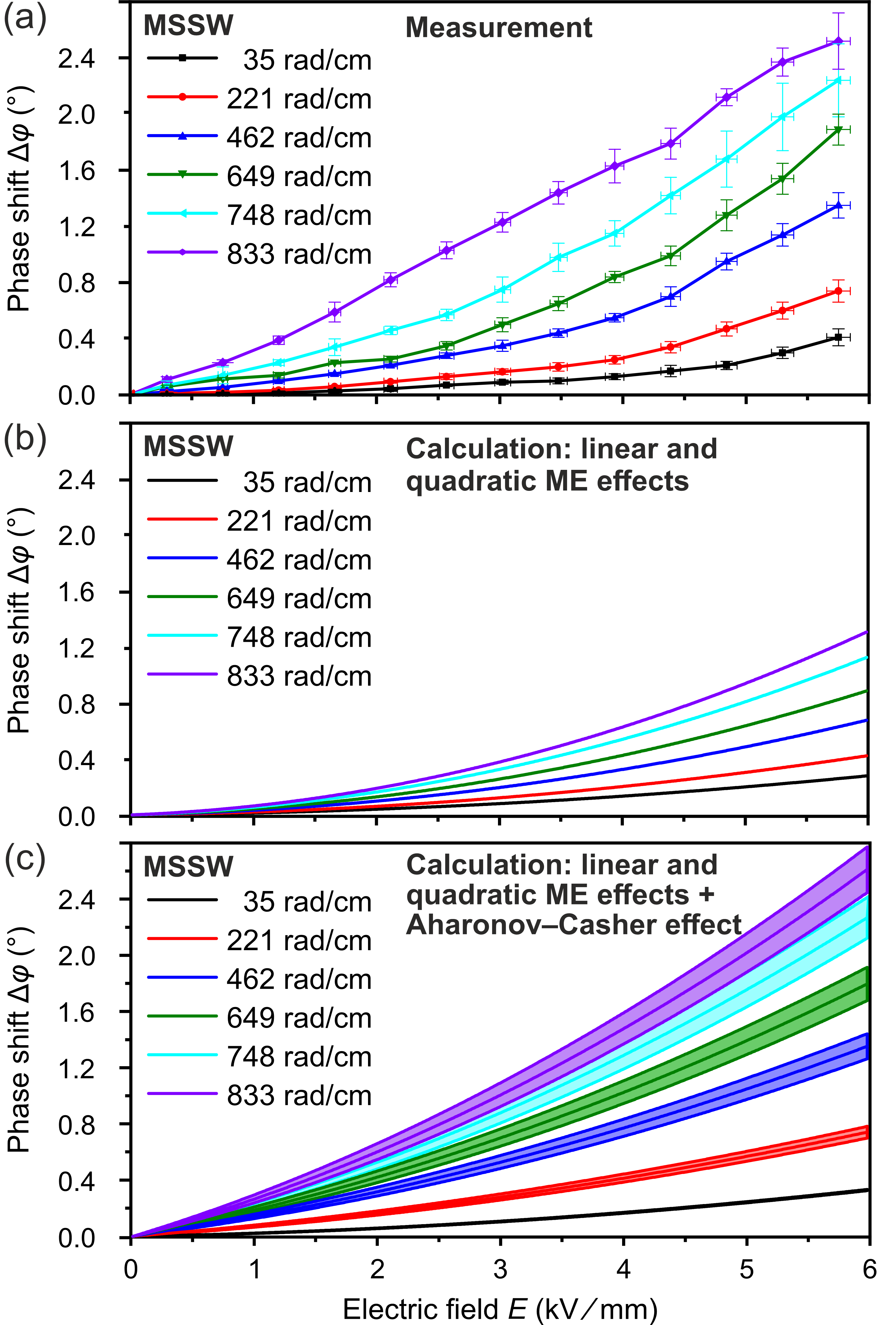}
		\caption{(a) Measurement of the MSSW phase shift vs. the applied electric field $E$ for six different wavevectors. (b) Calculations of the MSSW phase shift caused by quadratic and linear ME effects. The values of the ME coefficients are identical to those for the BVMSW calculations shown in Fig.\,\ref{f:2}b. (c) The MSSW phase shift $\Delta \varphi$ is calculated by adding the linear AC contribution to the phase changes caused by the linear and quadratic ME effects. The AC contribution is estimated by the coefficient $\kappa$ in the range $(6.16 - 8.46)\times 10^{-14}$\,kV\textsuperscript{-1}, which provides a $\Delta \varphi$ confidence band for each wavevector. The middle line in the bands refers to $\kappa=7.14 \times 10^{-14}$\,kV\textsuperscript{-1}, which is obtained by fitting the three experimental curves measured at the highest wavevectors.
		}
		\label{f:3}
\end{figure}

The phase-shift results are depicted in Fig.\,\ref{f:3}a. It becomes visible, that $\Delta\varphi$ strongly depends on $k$. While the maximum $\Delta\varphi$ for the smallest wavevector $k = 35$\,rad/cm is only about 0.4\textdegree, the maximum $\Delta\varphi$ for the largest $k = 833$\,rad/cm exceeds 2.5\textdegree. The $\Delta \varphi$ dependencies for other wavevectors are located between these curves in ascending order of $k$. Also, all phase curves, except the violet one, similar to the BVMSW case, have a clear quadratic dependence on the applied electric field for $E > 2$\,kV/mm. 
We can assume that this results from the quadratic ME effect in YIG. 

The YIG saturation magnetization $\mathbf{M}_\mathrm{s}$ changes quadratically with respect to $\mathbf{E}$ into a direction determined by a third rank tensor $\gamma_{ijk}$ \cite{Ascher1968, Rivera2009}. 
Unfortunately, the tensor coefficient $\gamma$\textsubscript{113} responsible for the magnetization change in the given geometry ($\mathbf{E} \perp \mathbf{M}_\mathrm{s}$ and $\mathbf{E} \parallel \langle 111 \rangle$) is unknown.
However, it is known that all these tensor coefficients are very small on the order of $\gamma \approx 10^{-13}$\,Am\textsuperscript{-1}V\textsuperscript{-2}m\textsuperscript{2} \cite{Cardwell1969}. This leads to only a tiny change in magnetization of about 6\,A/m for the maximal electric field in our experiment, while for YIG $M_\mathrm{s} = |\mathbf{M}_\mathrm{s}| \approx 140$\,kA/m at room temperature. Thus, it can be assumed that the change in the direction of magnetization caused by the electric field is negligible, and only a ${M}_\mathrm{s}$ change should be considered.

It can be also assumed that the YIG sample may also have a linear ME response in an electric field. In the ideal cubic YIG lattice, no linear response exists. However, in real YIG samples it is possible due to reduced crystal symmetry caused by the film shape, impurities, or a mismatch with the crystal lattice of the substrate \cite{Krichevtsov1989, Krichevtsov1994}. The coefficients for this effect are also unknown and are likely to be highly dependent on the quality and thickness of YIG films. 

The change in spin-wave energy caused by the Aha\-ro\-nov--Casher effect can be approximated as
\begin{equation}
	W_{k,\varphi}-W_{k,\varphi_{0}} \approx -C\mathbf{k}\cdot(\mathbf{\upmu_{m}}\times\mathbf{E})\,,
\end{equation}
where $C\approx J/(\hbar\mathrm{c}^2)$ with the exchange integral $J$ is a factor characterizing the spin-orbit interaction in vacuum, while $\mathbf{k}\cdot(\mathbf{\upmu_{m}}\times\mathbf{E})$ describes the energy of interaction between the magnon magnetic moment $\mathbf{\upmu_{m}}$ and an electric field $\mathbf{E}$ \cite{Krivoruchko2020}. 
Equation~(2) limits the possible spin-wave geometries available for the AC effect in our experiment since for BVMSW, when $\mathbf{k} \perp \mathbf{\upmu_{m}}\times\mathbf{E}$, the AC effect is expected to be very small, if not wholly vanishing. On the contrary, the MSSW geometry provides the maximum interaction with $E$ because $\mathbf{k}$ is parallel to $\mathbf{\upmu_{m}}\times\mathbf{E}$ in this case. This difference allowed us to determine the contribution of ME effects and distinguish the AC phase in the overall electrically induced phase shift of MSSW.  

First, we have fitted the linear and quadratic ME coefficients for the BVMSW modes, neglecting the AC effect. 
The least-squares fit minimizes the difference between our experimental data and the phase shift $\Delta \varphi_{_\mathrm{BVMSW}} = \Delta k_{_\mathrm{BVMSW}} l \times 180^\circ / \pi$ resulting from the ME-induced magnetization change in the BVMSW dispersion relation (see Eq.\,(5.97a) in Ref.\,\cite{Stancil2009})
	\begin{equation}\label{kBVMSW}
		\begin{aligned}
			&k_{_\mathrm{BVMSW}} = 2\sqrt{-(1+\chi)} \arctan\big(\sqrt{-(1+\chi)}\big)/d\,, \\
			&\chi = \omega_{_\textrm{H}} \omega_{_\textrm{M}} / \big( \omega^2_{_\textrm{H}} -\omega^2 \big)  \,, \\
			&\omega_{_\textrm{H}} = g\mu_0\textit{H}_\mathrm{ext} \,, \text{and}\; \omega_{_\textrm{M}}=g\mu_0\textit{M}_{\mathrm{s}}(E) \ .
		\end{aligned}
	\end{equation}
	Here, $g = 2\pi \times 28$\,GHz\,T\textsuperscript{-1},  $\omega = 2 \pi f$, $d$ is the film thickness, $l$ is the spin-wave propagation length in the electric field region, and $\textit{M}_{\mathrm{s}}(E) = \textit{M}_{\mathrm{s}}(0)+\alpha E+\gamma_{113} E^2 $.
	
These fits yielded the average values of the quadratic ME coefficient $\gamma_{113} = -1.46 \times 10^{-14}$\,Am\textsuperscript{-1}V\textsuperscript{-2}m\textsuperscript{2} and the linear ME coefficient $\alpha = -0.08 \times 10^{-7}$\,Am\textsuperscript{-1}V\textsuperscript{-1}m. The calculated phase-shift dependencies are shown in Fig.\,\ref{f:2}b. They agree with the measured phase shifts in shape and sign. Changing the wavevector in the range of values available in the experiment (48 -- 255\,rad/cm) has a minimal effect on the course of these curves. This is consistent with the fact that the difference in the experimental curves looks arbitrary and has no stable relationship with the wavevector magnitude.

Subsequently, we used the fitted coefficients to calculate the phase shifts of MSSWs using the following dispertion relation (see Eq.\,(5.111c) in Ref.\,\cite{Stancil2009}):
\begin{equation}\label{kMSSW}
k_{_\mathrm{MSSW}}=-\frac{1}{2d}\ln\Big[ 1 + \frac{4}{\omega^2_{_\textrm{M}}}\big(\omega_{_\textrm{H}}(\omega_{_\textrm{H}}+\omega_{_\textrm{M}}) - \omega^2 \big) \Big] \; . 
\end{equation}
The calculated phase shifts were compared with the experimental data to assess the contribution of the AC phase. This is possible because, in theory, the quadratic ME coefficients are the same in both cases (see Eq.\,6 in Ref.\,\cite{Ascher1968}). It is also assumed that the potential contribution of the linear effect of ME is at least very close in both cases. The latter assumption is related to the fact that the ME effect arises due to the reduction of crystal symmetry in the film. However, since the well-studied uniaxial magnetic anisotropy has the same origin \cite{Szymczak1979}, it can also be assumed that in the case of the linear ME effect, it acquires similar properties and can be described by identical coefficients for the MSSW and BVMSW geometries.

The calculated MSSW phase-shift dependencies are presented in Fig.\,\ref{f:3}b. Like the experimental data in Fig.\,\ref{f:3}a, they feature a positive phase shift. However, the calculated phase shifts are much smaller than the measured ones and are only about 40\% of them for the four largest wavevectors.  

It seems that in the case of MSSWs, there is indeed an additional contribution, which causes a more significant phase shift than the ME effects addressed. Thus, the next step is to consider the influence of the AC effect on MSSWs. In this paper, this is done empirically by adding a linear phase shift term that depends on the electric field and the spin-wave wavevector:
\begin{equation}\label{ACshift}
	\Delta\varphi_{_\mathrm{MSSW}} = \left[\Delta k^\mathrm{ME} + \kappa \, E \, k_{_\mathrm{MSSW}}(E) \right]l \times \frac{180^{^\circ}}{\pi}\ .
\end{equation}
Here, $\Delta k^\mathrm{ME}$ is the change in the MSSW wavevector $k_{_\mathrm{MSSW}}$ induced by the ME effects, and $\kappa$ is the empirical AC coefficient with unit $[\kappa]=$ kV\textsuperscript{-1}. Using the same ME coefficients and fitting $\kappa$ for each of the four largest wavevectors that show a significant deviation from the experimental results, we calculate the average value of $\kappa$, which is used in the new phase-shift calculations for all wavevector values. The results of these calculations are shown in Fig.\,\ref{f:3}c and can be compared with the experimental data in Fig.\,\ref{f:3}a. The average value of $\kappa$ is $7.14 \times 10^{-14}$\,kV\textsuperscript{-1}, but since there are variations in the fitting values for individual wavevectors, we decided to show the bands formed by the highest $\kappa=8.46 \times 10^{-14}$\,kV\textsuperscript{-1} and the lowest $\kappa=6.16 \times 10^{-14}$\,kV\textsuperscript{-1} fitting values. 
The obtained bands are in good agreement with the measured curves. 
It should be noted that the curves calculated for large $k$ are more sensitive to variations in the AC coefficient $\kappa$. This is evident from the width of the bands.

The good agreement of our estimations with experimental data indicates that MSSW magnons can accumulate an additional phase due to interaction with the electric field through the magnonic Aharonov--Casher effect. At the same time, additional experiments are needed to check the nonreciprocity of this difference upon inversion of electric or magnetic fields or the direction of spin-wave propagation.

In further experimental studies of this phenomenon, it seems advisable to use short-wave spin waves and microscopic interelectrode gaps. This will increase the AC phase and allow controlling it with a moderate external voltage. Notably, the quadratic ME effect in YIG films makes it unnecessary to apply excessively high electric fields, at which the phase shift caused by this effect significantly exceeds the AC phase. 
Success in this direction may lead to the observation of such manifestations of the magnonic Aharonov--Casher effect as persistent spin currents \cite{Cardwell1969, Nakata2015}, electrically induced refraction \cite{Krivoruchko2019}, non-reciprocity \cite{Savchenko2018}, anisotropy and caustics \cite{Krivoruchko2018_2, Savchenko2019} of exchange spin waves, and thus will allow the implementation of electrically controlled magnon optics and logic devices \cite{Liu2011}. 
This effect can also lead to magnon topological phases in insulating antiferromagnets \cite{Nakata2019} and to the magnonic quantum Hall effect \cite{Nakata2017_quantumHall}. It can be employed for the generation of twisted magnon beams carrying an electrically controllable orbital angular momentum \cite{Jia2019}. The implications of the magnonic Aharonov--Casher effect look particularly intriguing and exciting when the electric field is time-dependent and periodic \cite{Owerre2019}.

This research was funded by the European Research Council within the Advanced Grant No. 694709 “Super\-Magno\-nics.” The authors are grateful to R.V.\,Verba, V.S.\,L'vov, and M.\,Weiler for fruitful discussions.


%

\end{document}